\documentclass[letter,twocolumn,10pt]{article}

\usepackage{fullpage}
\usepackage{latexsym,amssymb, amsmath,subfigure,graphicx,url,psfrag}
\usepackage{algorithm}
\usepackage{algorithmic}
\usepackage{setspace}

\newcommand{\Put}{{\tt put}}
\newcommand{\Get}{{\tt get}}

\begin{document}

\title{A Peer-to-Peer Browsable File Index using a\\
Popularity Based Global Namespace}

\author{Thomas Jacobs and Aaron Harwood\\
\\
Department of Computer Science \& Software Engineering \\
University of Melbourne \\
Victoria, 3010, AUSTRALIA.}

\maketitle

\begin{abstract}
The distribution of files using decentralized, peer-to-peer (P2P) systems, has
significant advantages over centralized approaches. It is however more
difficult to settle on the best approach for file sharing. Most file sharing
systems are based on query string searches, leading to a relatively simple but
inefficient broadcast or to an efficient but relatively complicated index in a
structured environment. In this paper we use a browsable peer-to-peer file
index consisting of files which serve as directory nodes, interconnecting to
form a directory network. We implemented the system based on BitTorrent and
Kademlia. The directory network inherits all of the advantages of
decentralization and provides browsable, efficient searching.  To avoid
conflict between users in the P2P system while also imposing no additional
restrictions, we allow multiple versions of each directory node to
simultaneously exist -- using popularity as the basis for default browsing
behavior. Users can freely add files and directory nodes to the network. We
show, using a simulation of user behavior and file quality, that the
popularity based system consistently leads users to a high quality directory
network; above the average quality of user updates.
\end{abstract}

\section{Introduction}

Peer-to-peer (P2P) file sharing systems have steadily grown in
usage -- the Internet traffic generated in total by only seven P2P
file sharing systems was reported to have outgrown web traffic in 2002 and
increased to over half of all Internet traffic by the end of
2004~\cite{Dying,CacheLogic1}.  There are more than one hundred P2P file sharing
systems listed online and file sharing is the most widely
used application among other emerging applications of P2P including
\emph{Internet telephony}~\cite{voip}, \emph{instant messaging}~\cite{im},
\emph{grid computing}~\cite{grids},  and \emph{decentralized gaming}~\cite{dg}.

The P2P paradigm is loosely characterized by an application network in which a
significant proportion of the application's functionality is implemented by
\emph{peers} in a \emph{decentralized} way, rather than being implemented by
\emph{centralized} servers~\cite{1041681}. 
P2P \emph{file sharing} systems consist of program(s) that are used to create
and maintain P2P networks to facilitate the sharing of files between
users;  they allow users to designate a set of files from their PC's file
system to be shared and they allow users to download shared files from other
users of the P2P network. 

There are two key parts of a P2P file sharing system.  The first part is the
\emph{file distribution} system which provides the means to transmit files
between peers; it dictates how peers in the system should behave in order to
download and upload files. The second part is the \emph{file discovery} system
which is the means for users to find the files that are available on the P2P
network.  P2P file sharing systems typically provide the file discovery system
by maintaining some form of index of the files.  P2P file sharing systems
differ in how and where they implement these two parts. Some maintain the file
index in a centralized way, and others in a decentralized way; some indexes are
structured, i.e. provide efficient query processing, and some are unstructured
with inefficient query processing.  P2P file sharing systems implement the file
distribution system in a decentralized way; this is unequivocally the original
defining trait. 

In this paper we describe our experimental file sharing system, called
Localhost, that combines a directory node approach for file indexing with a
novel popularity based namespace. We show how the popularity based namespace
provides a way for decentralized maintenance to lead to a high quality
directory network.

\subsection{Query strings and browsing} 

Query string search is characterized as a process in which the user describes a
request by forming a query string that consists of one or more keywords and the
system presents a set of filenames that match or satisfy the query string.
The kind of index is transparent to the user.
Keyword search suffers from a vocabulary differences barrier, also referred to
as the semantic barrier (Nadis, 1996), between the publisher of the file and
the user wishing to download the file.  Keyword search is most suitable when
the user has an idea of what files they want from the system beforehand.
Keyword search is less suitable for presenting new things to users, as they
have to enter a specific query before being presented with a list of available
files.

Browsing is an alternate approach to find files. The user is presented with a
list of available files to select from. The index is seen by the user and
typically provides some categorization that allows the user to make a more
efficient selection. This approach does not suffer from the semantic barrier
because browsing presents a list of all available files.  However the user may
spend more time making a selection, especially when the list is large and the
index is flat, as compared to using query strings.

The majority of P2P file sharing systems use query string search.
Napster~\cite{napsterintro}, Gnutella-based~\cite{gnutella} systems,
eMule~\cite{eMule}, and KaZaA~\cite{Kazaa} currently use query string search as
their only means for finding files in their networks. Some P2P file sharing
systems allow the user to browse each individual peer's shared files.  However,
these systems do not support a browsable namespace that is global among all
peers because they do not directly provide a way of collaboratively organizing
files into a single, integrated, coherent categorical or hierarchical
structure.  Consequently, over 25 terabytes of files are fragmented across more
than 8000 individual listings, with each listing having its own way of
organizing its files~\cite{wayfinder}.

The Freenet system~\cite{journals/internet/ClarkeMHSW02} takes an unusual, but
necessary due to the anonymity property, approach of providing \emph{directory
nodes} that serve, like hyper-text documents, to point to other files in the
peer-to-peer network, thereby forming a browsable structure called a
\emph{directory network}, like the web. Leaves or end points of the directory
network are regular files.  In our work, we use this concept in a more general
way, transparently applying it to an existing P2P file distribution protocol.
Freenet does not allow different users of the system to write to the same name
in the namespace and this leads to a ``name race" situation where the first to
publish under a name will own that name. We consider the case when multiple
values of the same key can exist, and what kind of directory structure would
result. In this context, writing to a name in the namespace is synonymous with
sharing or adding a file to the network.  
 
\subsection{Adding files to the network}

A widely used method for adding files to the network is unstructured sharing,
where users designate a folder in their local file system and have all of its
contents shared.  The user shares files obliviously to what other files are
being shared and in many cases the files that are downloaded by the user from
another peer are also put into the user's local shared files folder. There is
no notion of a global namespace or index of all files in the network. 
  
Two major problems that occur in P2P file sharing systems that use unstructured
sharing are \emph{pollution} and \emph{poisoning}~\cite{poison}.  Pollution of
a P2P network is the accidental injection of unusable copies of files into the
network, by non-malicious users.  Poisoning is where a large number of
\emph{fake files} are deliberately injected into a P2P network by malicious
users or groups.  Fake files are specifically created by malicious users or
groups to seem like certain files, but consist of rubbish data or are unusable
in some way.  Both of these problems reduce the perceived availability of files
to users and reduce the usefulness of the system to users, because discovering
usable files is more difficult.  A study~\cite{Kazaa} found that a significant
proportion of files on the KaZaA network are unusable, due to poisoning and
pollution.  A number of P2P file sharing systems employ a \emph{file rating
system} in response to these problems.  File rating systems let users rate each
file's quality - the theory is that enough users will find the fake/unusable
files and rate them poorly, allowing other users to identify them before
downloading them.  These file rating systems have been shown to be largely
ineffective~\cite{Kazaa}.

The BitTorrent protocol specifies only file downloading, but a file sharing
system is nonetheless being used which is supported by the protocol.  Any user
can submit files to an index website, and the file is checked by the moderators
of the website before being added to the website's index.  If the file is found
to be fake or of unusable quality, it is not added to the index.  Although
pollution and poisoning levels are difficult to measure, sources indicate that
the the effective BitTorrent file sharing system is virtually pollution and
poisoning free because of this scheme~\cite{btindex}. While this system is
workable, it relies on a central server and it is difficult to decentralized the
moderation process, i.e. to allow all users to participate as moderators.

In our work we make use of a global namespace to store the directory network of
shared files. The global namespace is a set of names which are consistently
referred to by all peers in the network; each directory node and file has a
name in the namespace. A number of structured peer-to-peer protocols are
available that maintain a global namespace. The Kademlia protocol is used by
Azureus which we make use of in our implementation. While we considered a
number of existing shared access control methods, such as web based ownership
(the namespace is associated with IP addresses of users who can modify only
those parts of the name space that they control) and delegated authority (like
the domain name system where access is administered and delegated down through
a hierarchical authority), we proposed a new method based on popularity. In our
implementation, all users are allowed to submit their own version of the
content for a given name in the namespace; this method is appealing because the
users are still effectively acting obliviously to each other.  The system
naturally displays the most popular version to the user (in the case when the
user has no version viewing preference) when the content for that name is
requested, which is computed as the version that most users are currently
viewing. The user can optionally view all versions and make a different
selection at their discretion.

\subsection{Our contribution}

In our work, we apply the concept of a decentralized
directory network, using directory nodes that are transparently distributed by
an existing P2P file distribution protocol. We further show that a popularity
based namespace can be used to provide a way for decentralized maintenance to
lead to a high quality directory network. 

\section{The Localhost system}

\begin{figure*}[tb]
\psfrag{1. node name}{1. node name}
\psfrag{2. GET(H(nodename))}{2. $\Get(H(nodename))$}
\psfrag{3. list of versions}{3. list of versions}
\psfrag{4. GET(infohash)}{4. $\Get(H(version))$}
\psfrag{5. torrent file}{5. torrent file}
\psfrag{6. torrent file}{6. torrent file}
\psfrag{7. file content}{7. file content}
\psfrag{8. node display}{8. node display}
\begin{centering} 
\includegraphics[width=\linewidth]{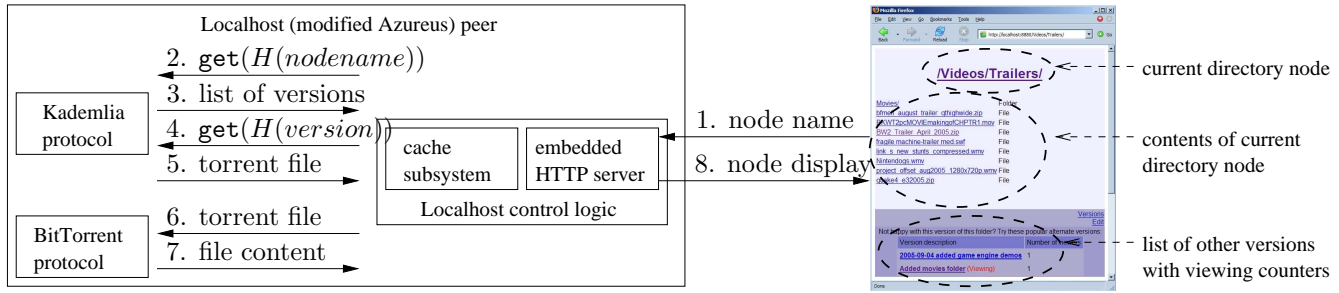}
\caption{Localhost system overview, showing example process and salient parts of the web browser
interface.}
\label{lgover}
\end{centering}
\end{figure*}

In this section we describe the details of the Localhost system, depicted in
Figure \ref{lgover}, to provide a context for the popularity based namespace
concept. At the highest level, the Localhost peer is a modification of the
Azureus peer; the modifications include additional data operations and an
embedded HTTP server to provide web browser based interaction. We use the term
\emph{Localhost distributed system} (LDS) to refer to the system that is
created by the interconnection of a number of Localhost peers. 

\subsection{Underlying protocols}

Our work builds from a number of technologies and in this section we abstract
the details that are sufficient to understand our modifications that were
applied to build the Localhost system. The technologies include the BitTorrent
file distribution protocol, the Kademlia distributed hash table (DHT), and
Azureus -- a user application which combines both BitTorrent and Kademlia.

\subsubsection{BitTorrent protocol}

The BitTorrent protocol is designed and used for P2P file
distribution~\cite{1064273,Cohen03}.  Following the BitTorrent protocol, a file,
$F=\{f_1, f_2,\dotsc\}$, is broken up into \emph{pieces} which are transmitted
between peers.  Piece size is usually between 32 kilobytes and 128
kilobytes.  A \emph{torrent file}, $T$, is used to publish a file or collection
of files and contains:
\begin{itemize}
\item the name(s) of the file(s), $F_1, F_2,\dotsc$,
\item the SHA-1 hash, $H(\bullet)$, of every piece of every file, 
\item the torrent file's \emph{infohash} which is the SHA-1 hash of
the concatenation of the file(s), $H(F_1F_2\dotsb)$ (or just $H(F)$ for
a single file) and
\item the web address for one or more \emph{trackers}.
\end{itemize}
The infohash is used to uniquely identify a torrent file.  The term
\emph{torrent} refers to the collection of file(s) that the torrent file was
created from.  From now on, without loss in generality, we will assume that the
torrent contains only a single file. A tracker is a server that maintains a
list of IP addresses of peers in the \emph{swarm}.  The swarm is the set of
peers currently involved in transmitting pieces of the file to each other. The
torrent file is distributed in full between users by some means external to the
BitTorrent peer, such as via web sites. The torrent file is input to the
BitTorrent protocol and is required for the protocol to download the file;
given the torrent file the peer contacts one or more of the listed trackers to
obtain the IP address and port numbers of other peers that are \emph{seeding}
the file. The user publishing the file acts as the initial \emph{seed} and
initially, there is one seed in the swarm.  As peers in the swarm obtain pieces
of the file, they become seeds for these pieces as well. 

\subsubsection{Kademlia protocol}

Kademlia~\cite{Kad} is one of many DHT based protocols, included among some of
the most well known such as Chord~\cite{chord}, CAN~\cite{CAN},
Pastry~\cite{Pastry}, and Tapestry~\cite{tapestry}. A DHT is global namespace
where each peer maintains some part of the space.  The two major DHT operations
that we consider are:
\begin{itemize}
\item $\Put(k, v)$ -- stores the data string $v$ under key $k$ in the DHT.
\item $v\leftarrow\Get(k)$ -- retrieves the data string $v$ from the
DHT that is stored under the key $k$. 
\end{itemize}

Note that some DHTs, including Kademlia, allow multiple data strings to be stored under,
and retrieved from, a single key. In this case, $v$ is a set of data strings.

DHT based systems operate in a completely decentralized way.  DHT protocols are
able to provide the two operations described above by making the peers form a
DHT \emph{overlay network}. The DHT overlay network is formed by each peer
maintaining a set of \emph{contacts}.  A contact is the \emph{peer ID} and IP
addresses of a remote peer in the DHT.  Each peer has a peer ID, which is a
number chosen from the namespace of keys.  The set of contacts each peer
maintains does not include every possible contact in the DHT.  The specific DHT
protocol used dictates which contacts each peer maintains.  Using these
contacts, DHT overlay networks such as Kademlia allow each peer to locate the
remote peer responsible for a certain key in $O(\log n)$ time.  Once the correct
peer has been located, the $\Put$ and $\Get$ operations can be done by
contacting that peer.

\subsubsection{Azureus application}

Azureus is a Java implementation of the BitTorrent protocol while allows a
number of files to be downloaded and seeded concurrently. From version 2.3,
Azureus also includes an implementation of the Kademlia protocol.  All Azureus
peers join the same DHT, by contacting a certain peer that is set up for the
purpose that aims to always be online.  Azureus uses the Kademlia DHT to
implement a feature called \emph{decentralized tracking}.  Decentralized
tracking is an optional replacement for BitTorrent trackers.  When decentralized tracking
is enabled, an Azureus peer executes the operation:
\[
\Put(H(F),(IP,port))
\]
for each file, $F$, that it is seeding; where $H(F)$ is the infohash for the
file and $(IP,port)$ is the peer's IP address and its BitTorrent protocol port
number. Multiple peers can store their $(IP,port)$ information at the same key.
Given $H(F)$ for a file, a peer can then execute $v\leftarrow\Get(H(F))$ to obtain a
list of other peers in the swarm for that file. The set of values in $v$ is
given to the BitTorrent protocol. This use of a DHT replaces the use of tracker
communication done by the standard BitTorrent protocol.

Version 2.3 of Azureus also introduces a torrent file download function
which allows torrent files to be downloaded from remote peers: 
$T\leftarrow\Get(H(F))$. The torrent file $T$ is then given to the BitTorrent protocol.

The Kademlia implementation in Azureus allows each peer to store a single value
only, under each key. The peer associates each stored value with the associated
peer's network address.  When a peer executes the sequence
\[
\Put(k, v_1),\Put(k,v_2),v\leftarrow \Get(k)\}
\]
then $v=v_2$ is the result.  In other words multiple peers can each store a
different value under the same key. A single peer can store different values
only under different keys.

\subsection{Localhost concepts}

In the following sections we describe the concepts that we proposed and
implemented using the previous technologies.

\subsubsection{Interpreting files as directory nodes}
\label{dirpage}
We adopted a directory node approach similar to the one taken by Freenet.  The
basic idea is to have the peer interpret some files as directory nodes, these
files contain an index of directory node names and/or file names.  The peer
displays this directory to the user and allows further selection.
Interestingly, because the directory nodes are distributed in a decentralized
way using the underlying file distribution protocol, the directory network
inherits this property, also becoming distributed in a decentralized way.
Another benefit of this approach for indexing files is that it can be applied
obliviously to the existing file distribution protocol.

Unlike web files which are served from centralized servers, P2P files are
served potentially from multiple peers.  Also, for files to be shared via the
directory network the peers must be able to add to or modify the directory
nodes. In our case, the use of a torrent file proved problematic because of its
use of hash functions. Consider a torrent file that contains, among other
things, the infohash of the file to be downloaded, $T=\{H(F)\}$; $T$ is
required to download $F$. It would be desirable to define a directory node by
inclusion of the torrent files for the files that are indexed by that directory
node. However, we cannot define a directory node to contain torrent files
because then for two directory nodes, $F_1$ and $F_2$, we would have a circular
reference, where $F_1=\{T_2\}$, $F_2=\{T_1\}$, $T_2=\{H(F_1)\}$ and
$T_1=\{H(F_2)\}$; hence $T_1=\{H(\{T_1\})\}$. This problem exists for any P2P
protocol that identifies files by using hash functions.  Let us restrict the
network structure to that of tree. The problem is then reduced to one of
efficiency, since for a chain of directory nodes, $F_1,F_2,\dotsc,F_l$, a
modification to $F_l$ requires a modification to $F_{l-1}$ and so on back to
$F_1$. Thus a root directory node would be modified for every modification to a
directory node beneath it in the tree.

Due to these relationships we separated the textual names of nodes, called node
names, from their infohash and make use of a two step process. A directory node
is defined as containing the node names of the nodes that it indexes. The
process of getting a node is numbered in Figure \ref{lgover}. A user request
for a node name generates a $\Get(H(nodename))$ which returns a list of
\emph{versions} of the node with that name. Versions are discussed in the next
section. Assuming a version is selected the peer executes a $\Get(H(version))$
to obtain a list of other peers that are seeding that file; any of these peers
can be contacted for the torrent file. The torrent file is then given to the
BitTorrent protocol to obtain the file content which is returned to the user.
If the file is a directory node then the directory structure is displayed and
the user continues to make selections.

\subsubsection{Node versions and popularity}

Systems that use a global namespace should specify how users can modify the
namespace in order to add the files that they wish to share; this immediately
poses the problems of shared access control when two or more users want to
modify content with the same name in the namespace. In our case the namespace
is the DHT space provide by Kademlia.

Interestingly the web uses a kind of global namespace, the set of uniform
resource locations, consisting of an IP address and file name; users can only
modify content with the names that they own on their local file system.
Outgoing connections can be easily made but incoming connections require
existing users to agree and to modify their existing files.  Because of this,
new content may not be linked to for some time if the publisher of that content
does not have agreements with existing publishers. We wanted to avoid this
situation for publishers of P2P files; peers should be able to effectively
operate independently of each other.

We considered the use of delegated authority, where the entire namespace is
initially owned by a single authority and permission to modify parts of it is
delegated on request; e.g. like the domain name system. We could implement this
approach by using a decentralized web of trust model. However, this approach
does not completely absolve peers from each other.

To adhere to the P2P paradigm, we proposed and implemented the notion of
versions. Each peer writes its own version of a given name in the namespace.
We make use of the DHT ability for multiple peers to each store a value at the
same name in the namespace. A version of a file is uniquely identified by the
infohash of that file. For the purpose of user selection, we also store a
textual description of each version along with its infohash. So for a given
file version, $F$, and a node name for that file, a peer executes 
\[
\Put(H(nodename),(description,H(F)))
\]
to register this version to the DHT. The peer of course must be seeding $F$.  A
$\Get(H(nodename))$ will proceed as discussed in the previous section to return
the list of all versions; and a selected version can be downloaded. Because the
list of versions could be as large as the number of peers, we use a download
time limit to download only a portion of the list relative to the speed of the
download. If the peer has not viewed that node before, then its viewing
preference is automatically set to the versions which is most popular (inferred
from the sample of versions collected in the download time limit). If the peer
has viewed that node before, then the viewing preference is whatever version of
that node the peer last viewed. A cache is used to maintain previously viewed
versions. This mechanism is the essential aspect of the popularity based
namespace.

Note that registering a version is effectively a ``preference" for that version
of the file or name in the namespace. As a consequence of the DHT allowing each
peer to write only one value for a given key, each peer can set a preference
for at most one version of each name in the namespace.

The Localhost peer provides appropriate web forms to the web browser for the
user to edit any currently viewed version of a directory or file node,
providing a new version to the system. When a peer downloads a version it also
registers the version, so that it contributes to the swarm of peers that share
that version.

\subsubsection{The user interface}

The dynamics of a popularity based system are in part influenced by the user
interface, and we have considered e.g. displaying the list of the most popular
versions, a list of all versions and a list of recently registered versions.
Discussion of the affect of the user interface on the system is beyond the
scope of this paper.

\section{Simulations of popularity based namespace}

In this section we show our simulation analysis of the Localhost system
with respect to use of a popularity based namespace. The goal of the analysis
is to understand the efficacy of the system, where we can loosely say that the
system is intended to provide peers with an ability to efficiently share files;
we further refine this to mean \emph{high quality} files. Broadly speaking, the
system should be \emph{malleable} in the sense that it can both admit a large
number of peer updates while remain coherent or stable in structure.

In our analysis, we generated a user model that represents user behavior.  For
this approach it is necessary to make assumptions about user behavior and also
to adopt a clear definition of \emph{quality} with respect to user updates,
e.g. the quality of a file or directory node update.  The main simplification
is that we consider the case when directory nodes can form connections only in
such a way that a tree network is formed, i.e.  new versions of directory nodes
can contain additional connections only to new directories or files, not to
existing directories or files. From now on we talk about the directory tree.
Our user model determines how a user traverses the directory tree, how they
make selections of possible versions and connections, how they choose to make
updates and what quality those updates have.

To assess the malleability of the system we applied our user model to a
starting directory tree containing only a few nodes and measured properties of
the resulting directory tree that is evolved by user updates. We measured
the ability for high quality updates to be ``seen" by other users in the
system and for high quality nodes to be viewed by the majority of users. In
this analysis we consider that updates are sequential and we consider the
evolution in terms of the update number.

\subsection{Directory tree and node quality}

The directory tree at time $t$ consists of a set of node versions, $V_t$ where
$v=v_{i,j}\in V_t$ is the $j$-th version for the $i$-th node, were
$i=1,2,\dotsc,n$ and $j=1,2,\dotsc,n_i$. The time $t$ represents the $t$-th
update to the tree.  The type of a node determines if it may contain
connections to other nodes or not; only directory nodes can contain
connections.  Each directory node has a set of connections to other nodes,
$E_t(v)\subset V_t$; where $E_t(v)=\{\}$ for all $t$ if $v$ is a file. File nodes
naturally represent leaves in the structure.

\begin{figure*}[t!]
\psfrag{p_update}{$p_{update}$}
\psfrag{p_add}{$p_{add}$}
\psfrag{p_file}{$p_{file}$}
\psfrag{R}{$\Rightarrow$}
\psfrag{no update}{no update}
\psfrag{delete a link}{delete a link}
\psfrag{add a directory}{add a directory}
\psfrag{add a file}{add a file}
\psfrag{v1,1}{$v_{1,1}$}
\psfrag{v2,1}{$v_{2,1}$}
\psfrag{v3,1}{$v_{3,1}$}
\psfrag{v4,1}{$v_{4,1}$}
\psfrag{v5,1}{$v_{5,1}$}
\psfrag{v1,2}{$v_{1,2}$}
\subfigure[Quality versus random value for various $s$]{\label{qualityfig}\includegraphics[width=0.24\linewidth]{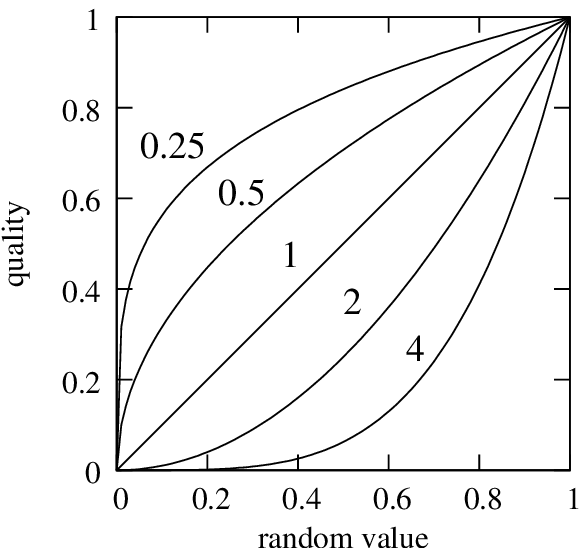}}
\subfigure[User choice versus random value for $k=4$ and various $\hat{\delta}$]{\label{choicefig}\includegraphics[width=0.24\linewidth]{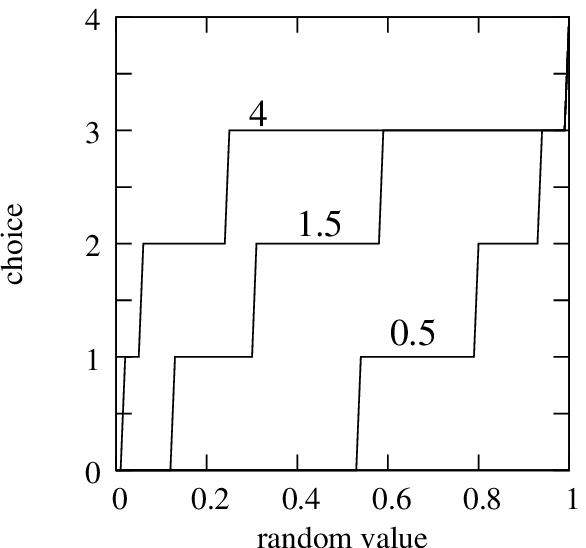}}
\subfigure[User decision process (top) and example update from the initial 
directory network (bottom)]{\label{userdecisions}\includegraphics[width=0.25\linewidth]{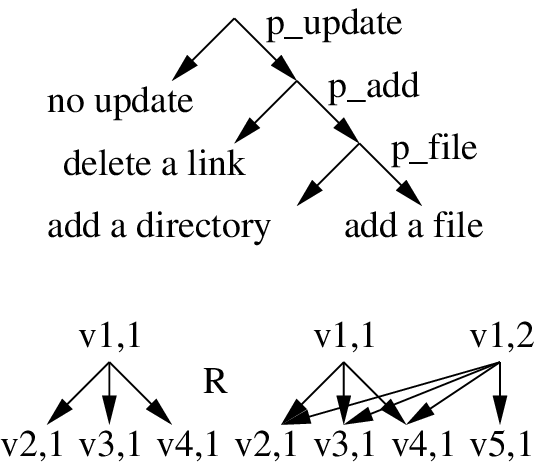}}
\subfigure[Example main tree of size 87 nodes after $10^5$ time steps]{\label{maintree}\includegraphics[width=0.24\linewidth,height=0.25\linewidth]{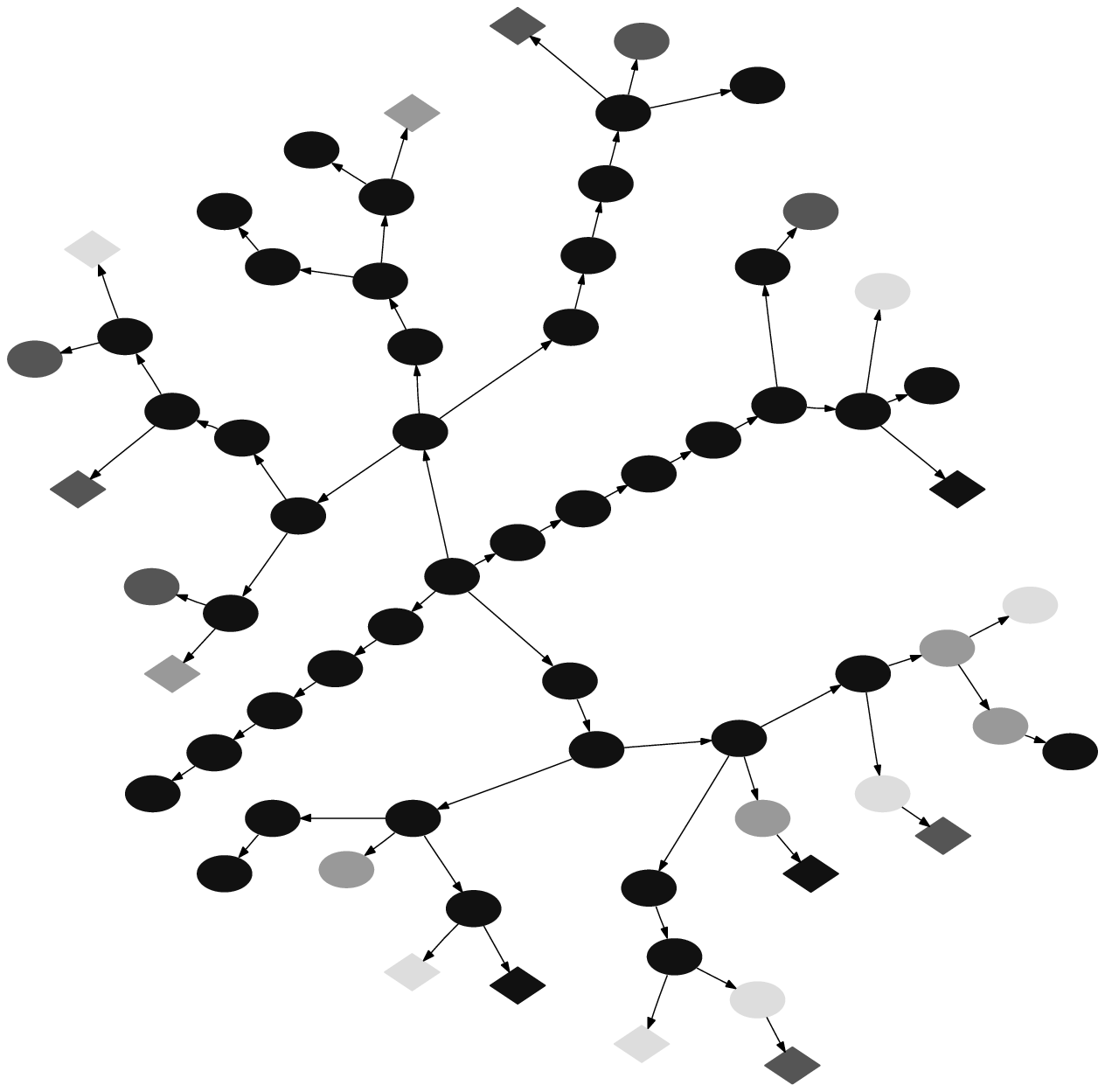}}
\caption{User model functions and examples}
\end{figure*}

We usually consider the case when the directory tree starts with
$V_0=\{v_{1,1}, v_{2,1},v_{3,1},v_{4,1}\}$,
$E_0(v_{1,1})=\{v_{2,1},v_{3,1},v_{4,1}\}$, all nodes are directories (there
are no files yet) and the leaf directories have no connections. Figure
\ref{userdecisions} (bottom) depicts the initial condition and an update, as
explained in the next section.

In~\cite{1066220} page quality is defined as the fraction of total users that
would like a page the first time they see the page; page quality is an
intrinsic property of the page. In our work, each node version $v$ has a
\emph{quality}, $Q_v\in[0,1)$. The quality determines the probability that a
user will continue to view that version rather than selecting a new version to
view; i.e. the probability of selecting a new version is $1-Q_v$ and this test
is made each time the user views that version. The quality of a node is
determined by a random variable and is set when the node is created by a user.
In our analysis, the quality is independent of all other nodes (including the
version it was derived from) and is independent of the user creating the
version. Thus, any changes to a node may induce an arbitrary increase or
decrease in quality. We use the cumulative distribution function
\[
\mathbb{P}\big[Q_v<q\big]=q^\frac{1}{s},
\]
where $s>0$ is a constant that describes the frequency of high quality updates
compared to low quality updates; if $s=1$ then all values of quality are
equally likely.  In this work, we use $\mathbf{p}$ to represent a source of
random numbers in the range $[0,1)$.
Thus, we choose the quality of a node $v_{i,j}$ using
\[
Q_{i,j}=\mathbf{p}_{i,j}^s.
\]
Figure
\ref{qualityfig} gives examples for various values of $s$. Note the expected number
of nodes with quality in $(a,b]$,
\[
\mathbb{E}\big[\lvert\{\,v\mid a\leq Q_v<b\,\}\rvert\big]=b^\frac{1}{s}-a^\frac{1}{s},
\]
and the average quality of the nodes:
\[
\mathbb{E}\big[Q_{i,j}\big]=\int_0^1p^{s}dp=\frac{1}{1+s}.
\]

\subsection{User behavior model}

Since we are interested only in the evolution of directory tree, we model the
user process as an update. In this work, the word user and peer is synonymous
since each peer is controlled by a unique user. The user starts from the root
of the tree and navigates to a leaf.  The user then chooses a node along the
navigated path, to make an update yielding a new version of that node. User
behavior is described by the parameters listed in Table \ref{user-actions}.
There are $N$ users.  The function $\gamma_{i,u}$ is set to the user $u$'s
version viewing preference for node $i$, it is undefined if the user $u$ has
not yet viewed node $i$. Each time that user $u$ traverses the tree, the
decision to update a node on the path is taken with probability $p_{update}$; a
new link is added with probability $p_{add}$ (otherwise an existing link is
deleted) and a new link points to a file with probability $p_{file}$ (otherwise
it points to a directory).  Note that an existing file can only be modified by
deleting the link to it in one traversal and then adding a new link in another
traversal.  Figure \ref{userdecisions} shows the decision process (top) and the
result of an update to the root node (creating version $v_{1,2}$) adding a new
link and hence a new node. 

Note that $p_{update}$ essentially allows us to model the \emph{update
frequency}, i.e. the fraction of time that is spent by users updating the
directory tree, rather than simply browsing and downloading. This is important
because the affects of popularity require time for users to group together on
the popular directory nodes, before being effective, and these affects in turn
affect the update outcomes.

\begin{table}
\caption{User state and actions}
\label{user-actions}
\begin{tabular}{l|p{0.75\linewidth}}\hline
notation & description \\ \hline
$N$ & the number of peers \\ \hline
$\gamma_{i,u}$ & the version of node $i$ that user $u$ is viewing \\ \hline
$p_{update}$ & the probability that a user makes an update for a given
traversal \\ \hline
$p_{add}$ & the probability that a user adds rather than deletes a link
in a directory \\ \hline
$p_{file}$ & the probability that a user adds a file rather than a directory
when adding a link \\ \hline
$p_{leave}$ & the probability that a user leaves the P2P network \\ \hline
\end{tabular}
\end{table}

Evolution of the directory tree is done by repeatedly calling Algorithm
\ref{update-process}, using a peer $u$ chosen uniformly at random from $N$,
increasing the time $t$ by 1 after each call; hence the number of ``time steps"
is the number of times that the algorithm has been called.

To model peer churn, we use the parameter $p_{leave}$; which is the
probability that a user leaves the network. When $u\in N$ is chosen for a traversal,
with probability $p_{leave}$ it will be ``reset". The reset erases all
popularity information in the directory tree (i.e. any versions that $u$ was
viewing). After a reset the (new) peer continues with the traversal. Thus, the
number of \emph{available} peers remains at a constant $N$, but peers
effectively come and go (once left, they do not return).

In Algorithm \ref{update-process}, user $u$ generates a path, $P$, starting
from the root and proceeding down to a leaf. The purpose of generating the path
is to model the behavior of a user who is browsing the structure to either
download a file or to make an update. It is not sufficient to simply pick at
random from the set of all nodes because the node probabilities are partially
determined by which users are viewing which nodes, the quality of the nodes,
and the connections from one directory node to another. Basically, the current
location of the peer is kept in $l=v_{i,j}$ and at each step, the peer checks
$Q_l$ to determine if a new version of node $i$ should be selected or deviated
to.  A selected version becomes the peer's viewing preference $\gamma_{i,u}$
for node $i$. The default version to view is a random version from the most
popular versions.  After traversing to a leaf, a decision is made whether to
update a node or not.

\begin{algorithm}
\caption{{\tt Traverse}(user $u$, time $t$)}
\label{update-process}
\begin{algorithmic}
\STATE $P \leftarrow \{\}$
\STATE $l=v_{1,j}\leftarrow \text{\tt Viewing}(1,u)$
\STATE $\delta \leftarrow \lvert E_t(l)\rvert$
\IF{$\mathbf{p}\geq Q_l$}
\STATE $l \leftarrow \text{\tt Select}(1)$
\ENDIF
\WHILE{$l$ is a directory node and $\lvert E_t(l)\rvert>0$}
\STATE $l=v_{i,j} \leftarrow \text{\tt Viewing}\big(\text{\tt Random}(E_t(l)),u\big)$
\STATE $\delta \leftarrow \delta + \lvert E_t(l)\rvert$
\IF{$\mathbf{p}\geq Q_l$}
\STATE $l \leftarrow \text{\tt Select}(i)$
\ENDIF
\IF{$l$ is not a file}
\STATE $P\leftarrow P\cup\{l\}$
\ENDIF
\ENDWHILE
\STATE $\hat{\delta} \leftarrow \delta/\lvert P\rvert$
\STATE $c \leftarrow$ the $C(\hat{\delta},\lvert P\rvert)$-th entry in $P$
\IF{$\mathbf{p}<p_{update}$}
\STATE $\text{\tt Update}(c,u)$
\ENDIF
\end{algorithmic}
\end{algorithm}

The model to determine which node in $P$ to update also requires consideration.
Choosing uniformly at random would cause excessive updates to the root node
which is unrealistic. We model the user choice by computing an estimate size of
the total directory tree, based on the outgoing degree of directory nodes along
$P$ and the total length of the path, and then generating a staircase
probability distribution that provides an approximate uniform random
distribution over all accessible nodes. This model says that users are more
likely to make updates towards the leaves of the tree rather than towards the
root.

From Algorithm \ref{update-process}, $\delta$ is the total degree of the nodes
in the path (not including node versions that were deviated from because of a
quality decision, and not including the last node if it is a file). Then
$\hat{\delta}=\delta/k$ where $k=\lvert P \rvert$ and the choice of which node
to update is given by:
\begin{equation}
C(\hat{\delta},k)=\big\lfloor\log_{\hat{\delta}}(1+(\hat{\delta}^{k}-1)\mathbf{p})\big\rfloor
\end{equation}
where $\mathbf{p}$ is chosen uniformly at random in $[0,1)$. Examples are shown
in Figure \ref{choicefig}; the choice function is in an integer in $\{0,1,\dotsc,k-1\}$
and is never equal to $k$ because $\mathbf{p}$ is never 1. Note that:
\[
\lim_{\hat{\delta}\rightarrow 1}C(\hat{\delta},k)=\lfloor \mathbf{p}\,k\rfloor
\]
which is the case when the tree appears to be a linear list.

To make random selections with probability proportional to the number of
viewers of a version, and to select randomly among the most popular versions,
we let $\lambda(v_{i,j})=\big\lvert\{\,u\mid j=\gamma_{i,u}\,\}\big\rvert$, and
$\lambda_{max}(i)=\max_j\{\lambda(v_{i,j})\}$. The function {\tt Random}(set
$X$) returns an element $x\in X$, chosen uniformly at random. Algorithms
\ref{viewing}, \ref{select} and \ref{update} are called by Algorithm
\ref{update-process}.

\begin{algorithm}
\caption{{\tt Viewing}(node index $i$, peer $u$)}
\label{viewing}
\begin{algorithmic}
\IF{$j=\gamma_{i,u}$ is undefined}
\STATE $j=\gamma(i,u) \leftarrow \text{\tt Random}\big(\{\,v_{i,j}\mid \lambda(v_{i,j})=\lambda_{max}(i)\,\}\big)$
\ENDIF
Return $v_{i,j}$ 
\end{algorithmic}
\end{algorithm}

\begin{algorithm}
\caption{{\tt Select}(node index $i$)}
\label{select}
\begin{algorithmic}
\STATE $v\leftarrow v_{i,j}$ with probability $\lambda(v_{i,j})/\sum_j \lambda(v_{i,j})$
Return $v$ 
\end{algorithmic}
\end{algorithm}

\begin{algorithm}
\caption{{\tt Update}(node $v_{i,j}$, user $u$)}
\label{update}
\begin{algorithmic}
\STATE $j^\prime \leftarrow n_i\leftarrow n_i+1$
\STATE $E_{t}(v_{i,j^\prime}) \leftarrow E_t(v_{i,j})$
\STATE $Q_{i,j^\prime}\leftarrow \mathbf{p}_{i,j^\prime}^s$
\IF{$\mathbf{p}>p_{add}$ and $|E_t(v_{i,j^\prime})|>0$} 
\STATE delete the connection $\text{\tt Random}(E_t(v_{i,j^\prime}))$
\ELSIF{$\mathbf{p}>p_{file}$}
\STATE $n \leftarrow n+1$
\STATE node $v_{n,1}$ becomes a directory node 
\STATE $E_t(v_{i,j^\prime})\leftarrow E_t(v_{i,j^\prime}) \cup \{v_{n,1}\}$
\STATE $Q_{n,1}\leftarrow \mathbf{p}_{n,1}^s$
\ELSE
\STATE $n \leftarrow n+1$
\STATE node $v_{n,1}$ becomes a file node 
\STATE $E_t(v_{i,j^\prime})\leftarrow E_t(v_{i,j^\prime}) \cup \{v_{n,1}\}$
\STATE $Q_{n,1}\leftarrow \mathbf{p}_{n,1}^s$
\ENDIF
\end{algorithmic}
\end{algorithm}

\subsection{Simulation control parameters}

We used the control parameters $N=100$, $s=1.0$, $p_{add}=0.75$,
$p_{file}=0.5$, $p_{update}=0.5$, $p_{leave}=0.0$. The initial directory
tree is given in Figure \ref{userdecisions} (bottom left) and is viewed by node
0, all nodes are directories with quality 0.5. We ran all the simulations until
time $t=10^5$ and the results are the average of ten realizations. The control parameters
correspond to a fixed number of dedicated peers that are vigorously updating the
directory tree, adding new nodes equally likely to be files or directories.

When examining the results we often consider the \emph{main tree}, which we
define as the tree that a new peer, having no initial viewing preferences,
would browse. This tree is computed by tracing the most popular paths. In the
case that more than one such tree exists then we pick at random. An example
main tree is shown in Figure \ref{maintree}, for the control parameters.
Ellipses are directories, diamonds are nodes and there are 4 shades of
grey, from dark grey which indicates quality $>0.75$ to light grey which
indicates quality $<0.25$. In the example notice that quality is generally
higher towards the root, because those nodes are visited more often leading to
a better efficacy of the popularity effect.

The \emph{average quality} of the main tree is defined as the average of the
quality of all nodes in the main tree. The outcome is good if the average
quality exceeds $1/(1+s)$ and bad if it does not.

We independently varied each of the parameters and report the most interesting
results in the following sections.

\begin{figure*}[t!]
\subfigure[Average nodes in main tree versus time]{\label{MainTreeNodesVersusUpdateFrequency}\includegraphics[width=0.24\linewidth]{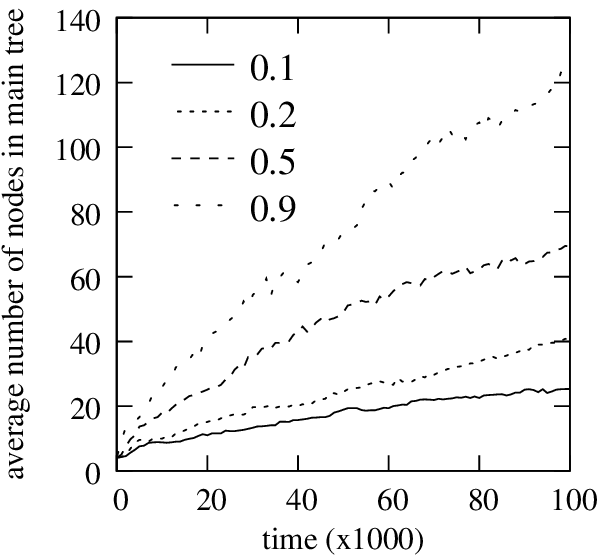}}
\subfigure[Node frequency versus degree at $t=10^5$]{\label{NodeFrequencyVersusDegreeForUF}\includegraphics[width=0.24\linewidth]{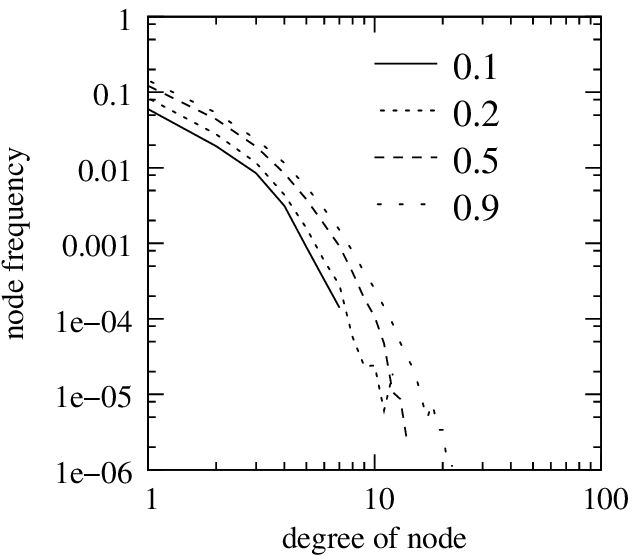}}
\subfigure[Node frequency versus numbers of viewers at $t=10^5$]{\label{NodeFrequencyVersusUpdateFrequency}\includegraphics[width=0.24\linewidth]{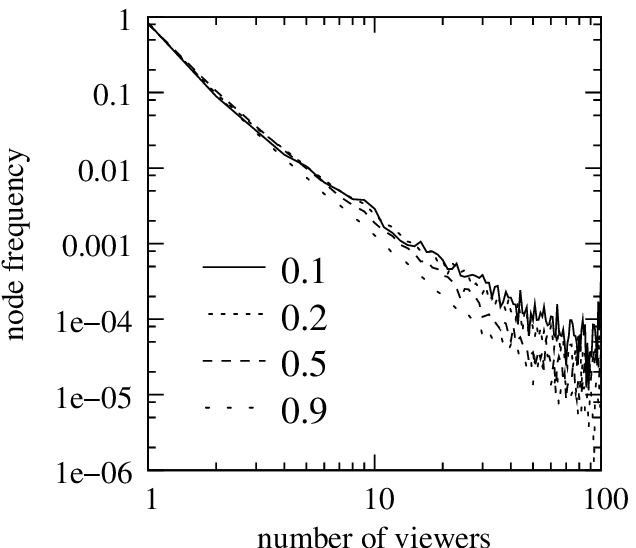}}
\subfigure[Viewers per node versus quality of node at $t=10^5$]{\label{ViewersPerNodeVersusQualityForUF}\includegraphics[width=0.24\linewidth]{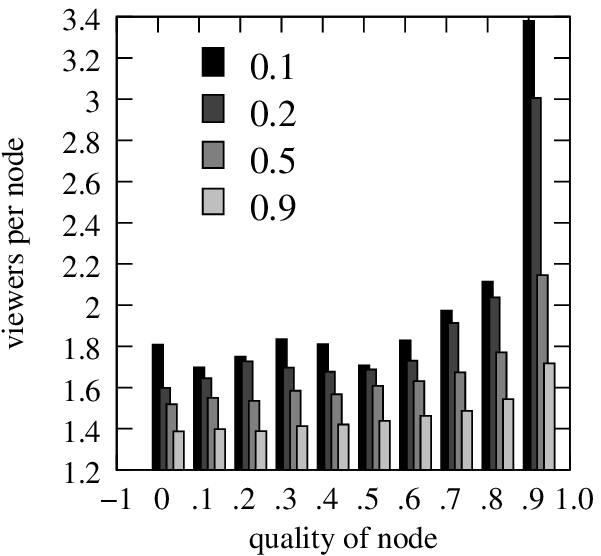}}
\subfigure[Average quality of the main tree versus time]{\label{MainTreeQualityVersusTimeForUF}\includegraphics[width=0.24\linewidth]{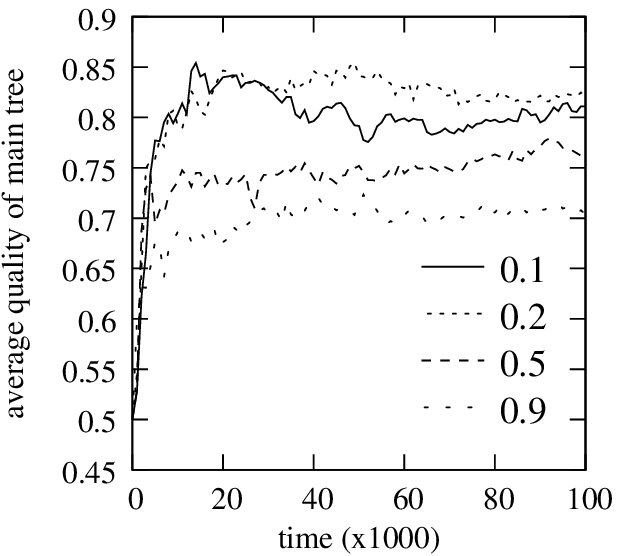}}
\subfigure[Number of nodes that reach a majority versus quality]{\label{NumNodesReachMajorityForUF}\includegraphics[width=0.24\linewidth]{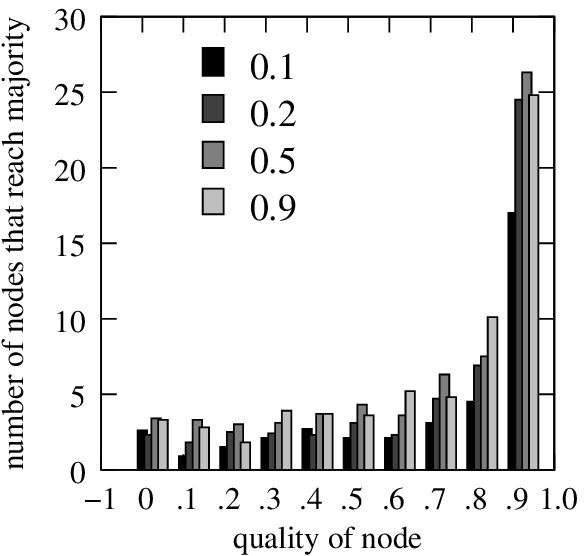}}
\subfigure[Average time to reach a majority versus quality]{\label{TimeTakenForUF}\includegraphics[width=0.24\linewidth]{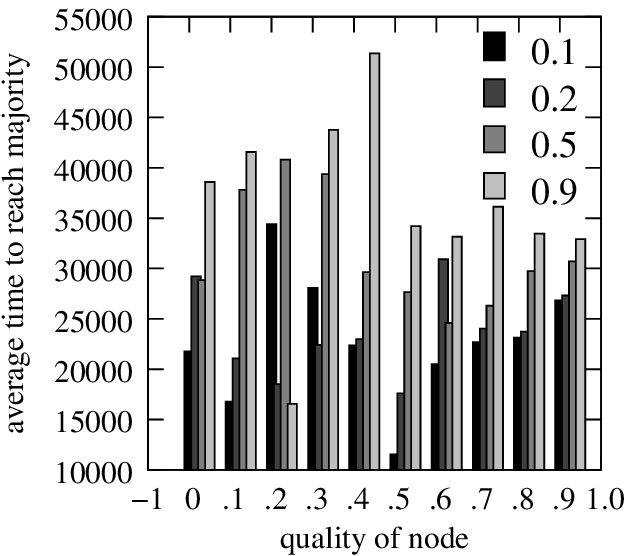}} 
\subfigure[Average quality of the main tree versus time with varying $s$]{\label{AverageQualityVersusTimeForS}\includegraphics[width=0.24\linewidth]{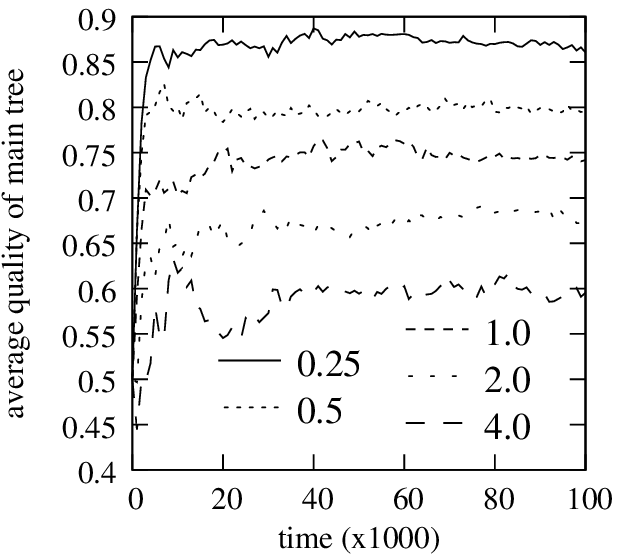}}
\caption{Simulation results, (a) to (g) with $p_{update}=\{0.1,0.2,0.5,0.9\}$, (h) with $s=\{0.25,0.5,1.0,2,4\}$}
\end{figure*}

\subsection{Update frequency}

The update frequency is given by $p_{update}$ and determines the proportion of
traversals of the directory tree that result in updates. A low update frequency
means that users are browsing the tree more often than updating, and vice
versa. Clearly the total nodes in the system increases to become roughly
$10^5p_{update}$ (e.g. with $p_{update}=0.9$ it increased to nearly 90,000;
nodes with no viewers are not counted), however the average nodes in the main
tree is around $0.1-0.2\%$ of the total nodes, as shown in Figure
\ref{MainTreeNodesVersusUpdateFrequency}.  Figure
\ref{NodeFrequencyVersusDegreeForUF} provides the total node frequency versus
degree.  A smaller update frequency leads to nodes of smaller degree.

The node frequency versus number of
viewers, Figure \ref{NodeFrequencyVersusUpdateFrequency}, shows that almost
all of the nodes have only 1 viewer (the creator of that node) and that this
distribution becomes more negatively sloped as the update frequency increases.
This is because new versions, regardless of quality, are selected with
probability $1/N$ and if the new version is contained in a tree outside of the
main tree then its probability of selection is further reduced. At the same
time, Figure \ref{ViewersPerNodeVersusQualityForUF} shows that the number
of viewers per high quality node ($Q\in[0.9,1)$) is almost twice as much
as that for a low quality node ($Q\in[0,0.1)$) when $p_{update}=0.1$, but
this difference is reduced as $p_{update}$ increases. Clearly, a small update
frequency allows the popularity of high quality nodes to become more distinguished
than low quality nodes.

Figure \ref{MainTreeQualityVersusTimeForUF} shows the average quality of the
nodes in the main tree versus time. A small update frequency leads to a
significantly better average quality, though there is no difference from
$p_{update}=0.2$ to $0.1$; so reducing the update frequency further than this
does not help. Note that the average quality of an update is $1/(1+s)=0.5$; so
the system is working well to improve the average quality of files found.
Compare Figure \ref{MainTreeQualityVersusTimeForUF} to Figure
\ref{AverageQualityVersusTimeForS}, which shows the average quality of the main
tree when only $s$ varies from $0.25$ to $4$. In all cases the average quality
of the main tree is above the mean quality, $1/(1+s)$, over all nodes. 

\begin{figure*}[t!]
\subfigure[Node frequency versus degree at $t=10^5$]{\label{NodeFrequencyVersusDegreeForPFile}\includegraphics[width=0.24\linewidth]{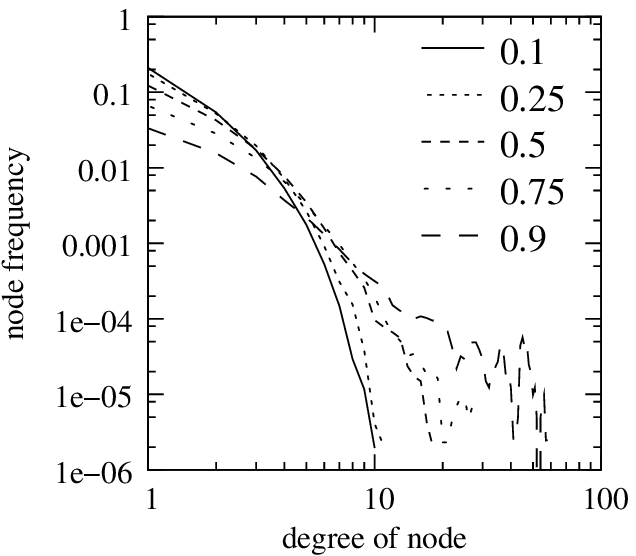}}
\subfigure[Viewers per node versus quality of node at $t=10^5$]{\label{ViewersPerNodeVersusQualityForPFile}\includegraphics[width=0.24\linewidth]{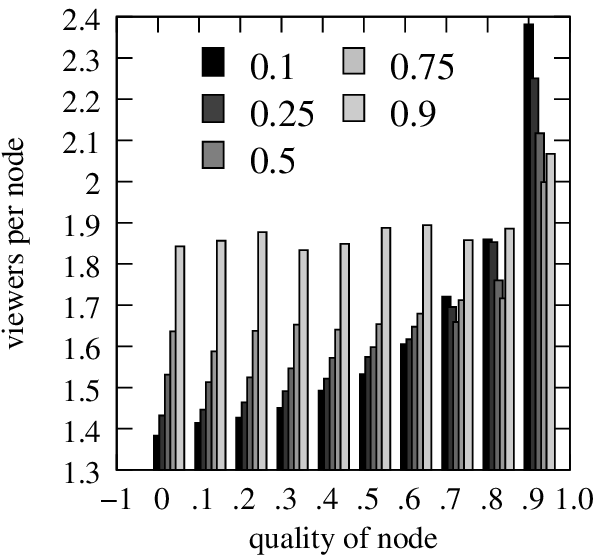}}
\subfigure[Average quality of the main tree versus time]{\label{MainTreeQualityVersusTimeForPeers}\includegraphics[width=0.24\linewidth]{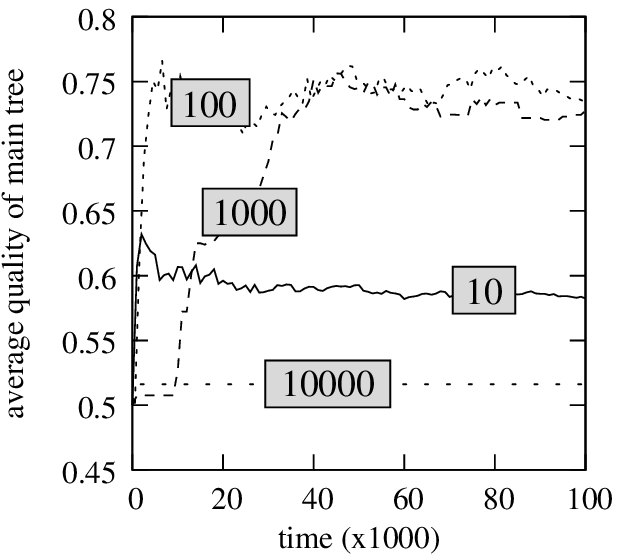}}
\subfigure[Node frequency versus degree at $t=10^5$]{\label{NodeFrequencyVersusDegreeForPeers}\includegraphics[width=0.24\linewidth]{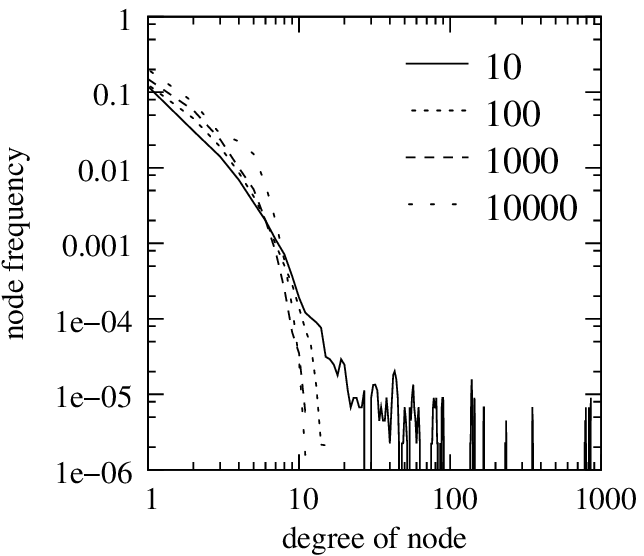}}
\subfigure[Node frequency versus number of viewers at $t=10^5$]{\label{NodeFrequencyVersusNumViewersForS}\includegraphics[width=0.24\linewidth]{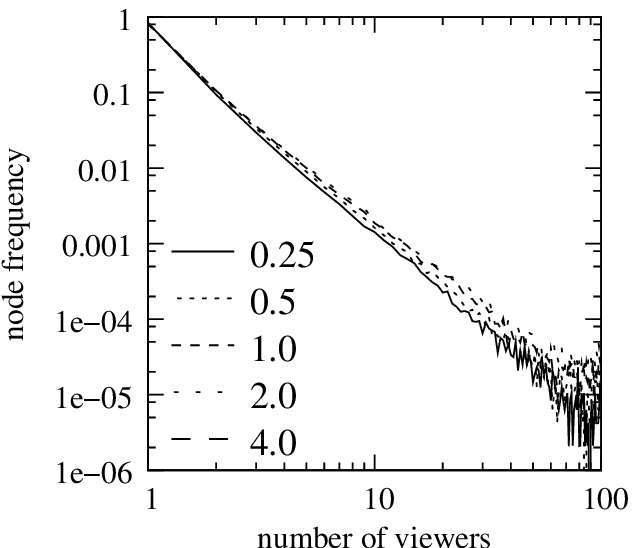}}
\subfigure[Number of nodes that reach a majority versus quality]{\label{NumNodesReachMajorityForS}\includegraphics[width=0.24\linewidth]{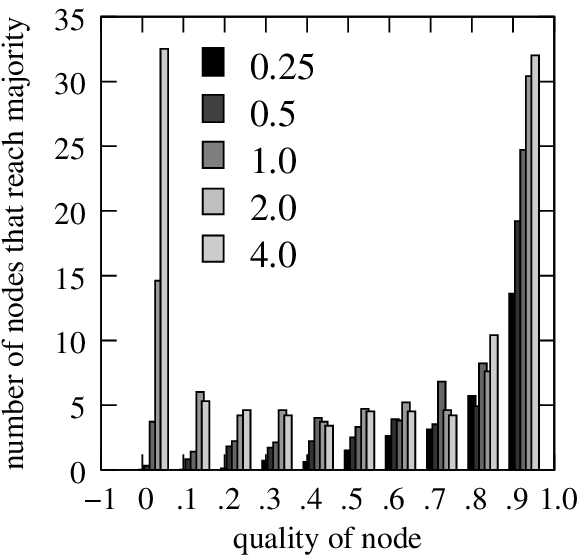}}
\subfigure[Average quality of the main tree versus time with $s=4$]{\label{MainTreeQualityVersusTimeForSUG}\includegraphics[width=0.24\linewidth]{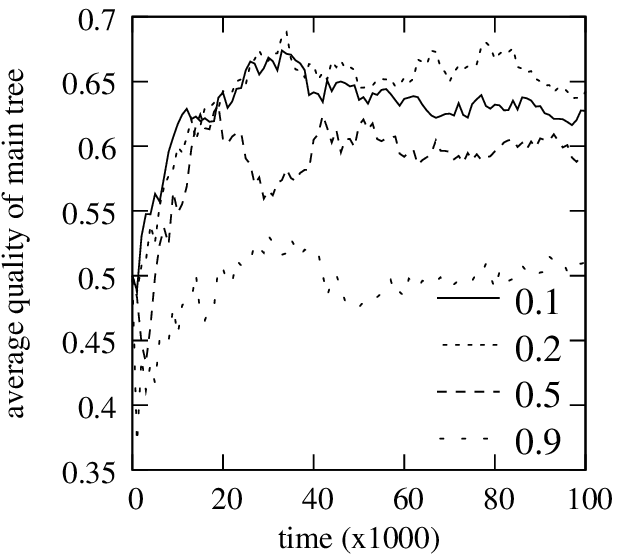}}
\subfigure[Visual depiction of main tree for $N=10$ at $t=10^5$, the dark center
is the result of excessively large degree]{\label{VisualForPeers}\includegraphics[width=0.24\linewidth]{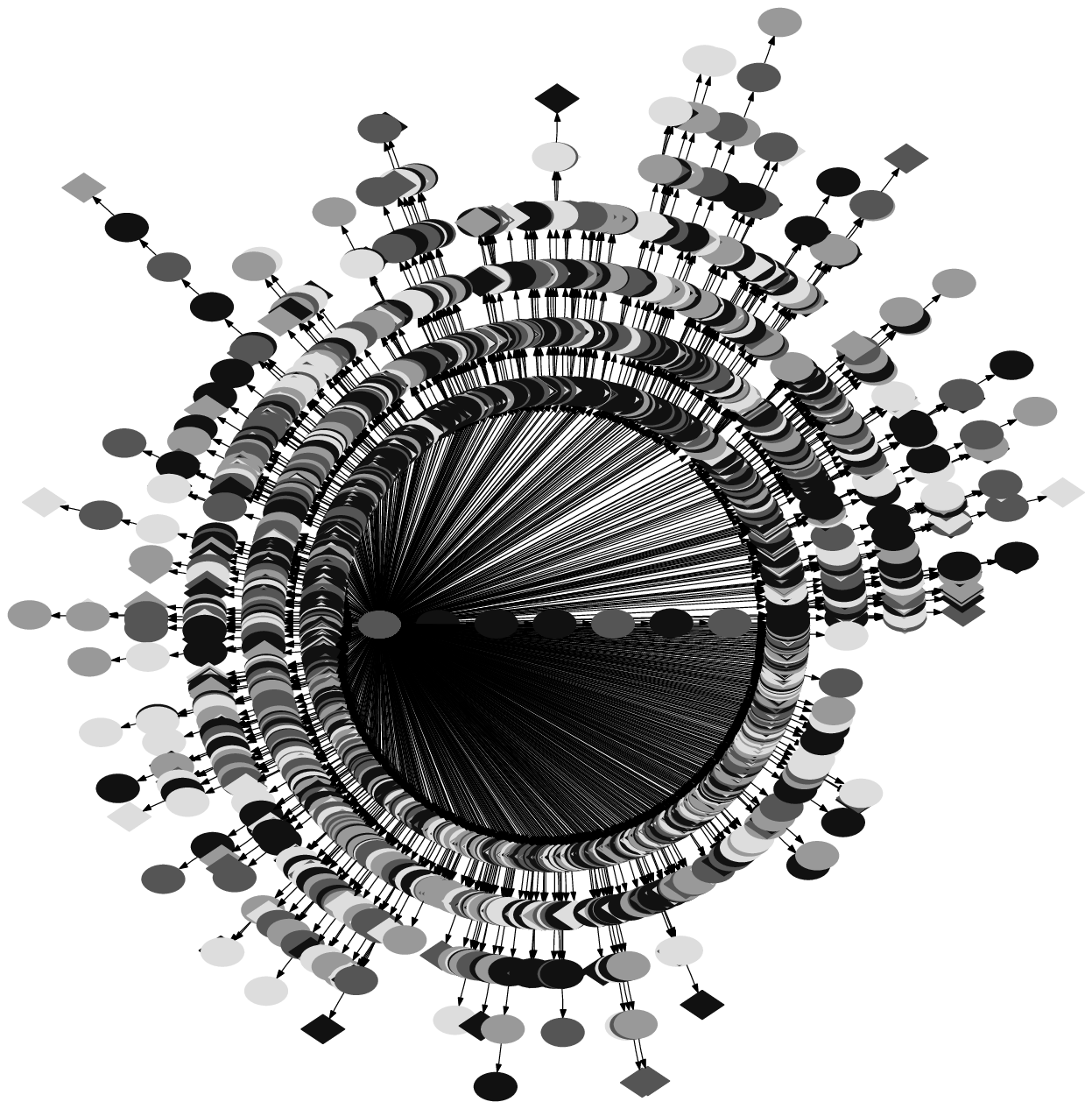}}
\caption{Simulation results, (a) - (b) for $p_{file}=\{0.1,0.25,0.5,0.75,0.9\}$,
(c) - (d) for $N=\{10,100,1000,10000\}$, (e) - (f) for $s=\{0.25,0.5,1.0,2.0,4.0\}$,
(g) for $s=4$ and $p_{update}=\{0.1,0.2,0.5,0.9\}$, (h) for $N=10$}

\end{figure*}

Continuing, Figure \ref{NumNodesReachMajorityForUF} shows the number of nodes
that reach a majority (more than half the users) and Figure
\ref{TimeTakenForUF} shows the average time it takes from the time of creation,
for a node of given quality to reach a majority. Interestingly there is no
apparent trend in Figure \ref{TimeTakenForUF}. This is due to the fact that
users select a version to view based only on popularity and not node quality.
Consider a low quality node that gains moderate popularity; while, with high
probability, users migrate away from the low quality node, users are likely to
choose the node of moderate popularity over a high quality node that has little
popularity at that time. A low quality node could gain mild popularity via
random fluctuations. It could also quickly gain high popularity if it is the
only node on a path. Even if a new version is created with high quality, the
already popular low quality node remains popular for some time.

\subsection{Add frequency}

We set the add frequency, $p_{add}$, to be $\geq 0.5$ so that the growth of the
directory tree was positive. As $p_{add}\rightarrow 0.5$ the average quality of
the main node increases, similarly to reducing $p_{update}$. However the total
number of nodes decreases because in some cases a deletion event is
ineffectual, e.g. a node cannot be deleted outright in the sense of removing it
from the system, it can only be contended with a new version; a node can be
effectively removed from the main tree by creating a high quality version of
the parent node that does not contain the child connection. 

\subsection{File/Directory frequency}

The file/directory frequency, $p_{file}$, determines how often files are added
as opposed to directories. As files become more likely, the number of
directories decreases. This naturally increases the degree of the directories
as shown in Figure \ref{NodeFrequencyVersusDegreeForPFile}; furthermore it
tends to push the degree distribution towards a power law.  While this leads
only to a slight decrease in average quality of the main tree, the number of
low quality nodes that reach a majority increases to the point that there is
very little distinction between low quality nodes and high quality nodes; the
number of viewers per low quality node becomes roughly equal to the number of
viewers per high quality node as shown in Figure
\ref{ViewersPerNodeVersusQualityForPFile}.

\subsection{Number of peers and churn}

Increasing the number of peers shows an interesting outcome which is seen in
Figure \ref{MainTreeQualityVersusTimeForPeers}, the quality of the main tree
versus time. For $N=10$ the average quality falls to a value less than that for
$N=100$.  For $N=1000$ it rises to match the case when $N=100$ though more
slowly. For $N=10000$ it sits slightly above $0.5$ which is the initial quality
of the initial nodes; in a separate simulation over twice the time interval we
observe the average quality to rise to almost 0.6, hence as $N$ becomes large
it takes longer for the main tree to grow.  For a small number of users, a node
can quickly become popular, but it does not necessarily stay popular for very
long. For a large number of users, it takes longer for a node to become
popular, and popular nodes remain popular for longer. Hence the average size of
the main tree grows quickly for $N=10$ (it reaches over $1500$ nodes, with an
example in Figure \ref{VisualForPeers}) and grows
slower as $N$ increases (reaching less than 100 for $N=100$, and no significant
growth showing for $N=10000$). The rapid growth for small $N$ is also reflected
in the node frequency versus degree, Figure
\ref{NodeFrequencyVersusDegreeForPeers}.  However, unless there are a
sufficient number of users, the affect of popularity on finding (and keeping)
high quality nodes is less and so the average quality of the main tree is less.

Changes in churn, $p_{leave}$, between $0.1$ and  $0.9$ had little to no
affect on any of the measures. This is because popular nodes that loose a
viewer due to churn are likely to receiver a replacement viewer. High churn
rates do lead to a small increase in the average size of the main tree.

\subsection{Quality parameter}

Varying the quality parameters, $s$, has the obvious outcome of varying the
average quality of nodes found in the system and the various is shown in Figure
\ref{AverageQualityVersusTimeForS}.  As the average quality of nodes decreases,
users are more likely to search for a better quality version. Since a better
quality version is of lower frequency they attract and keep larger numbers of
users; hence we see a positive increase of slope with decreasing $s$ in Figure
\ref{NodeFrequencyVersusNumViewersForS}. We also see an increase in the number
of low quality nodes that reach a majority, along with an increase in the
number of high quality nodes that reach a majority, with an increase in $s$,
shown in Figure \ref{NumNodesReachMajorityForS}. There are more low quality
nodes and so there is a larger that become popular as the users search for high
quality nodes. There are less high quality nodes and so less competition and
hence more high quality nodes can a majority.

When $s=4$ and the update frequency varies from low to high then the average
quality of the main tree reaches as high as 0.65 (compared to average of all
updates which is 0.25).
Compared to Figure \ref{MainTreeQualityVersusTimeForUF}, the peak quality of
the main tree has dropped from roughly 0.85 to 0.7, while the average quality
dropped from 0.5 to 0.25.

\subsection{Summary}

In all cases that we have observed the quality of the main tree is consistently
above that of the average quality, even when node updates are relatively
frequent and different numbers of peers are using the system with high churn
rates. However the size of the main tree is typically less than 1\% of the
total number of nodes because many of the updates are never viewed more than
once. The growth of the main tree is significantly affected by the number of
users. The tree grows rapidly for a small number of users and takes a long time
to grow for a large number of users. However when the number of users is larger
then the average quality of the main tree increases. The natural search
process, as lead by popularity, causes even low quality files to become popular
at times; a low quality file can become popular just as quickly as a high
quality file because quality is not known by a user until the node is viewed by
the user. However the high quality files sustain popularity for a longer time.

\section{Related work}

The conventional web system allows users to post files and connect those files
to files posted by other users. The web system, including clients, can be
considered as a centralized directory network in the sense that web clients do
not participate in the distribution of web files; also the failure of a single
(popular) web server, e.g.  a web directory site, may cause significantly more
harm than the failure of most other web servers.  Freenet is an example of
providing a decentralized directory network. However the Freenet system has
anonymity requirements that place restrictions on how that directory network
can be used by peers. 

The \emph{Open Directory Project (ODP)}~\cite{opendirproject} is a human-edited
directory structure which indexes websites.  It indexes websites in a
hierarchical structure, and is itself a website.  The nodes in the hierarchical
structure are categories, and the leaves are website links.  The top level
nodes are broad categories, such as Arts, Business, Computers, and News.  The
ODP is constructed and maintained by a global community of volunteer editors.

\emph{Wikipedia}~\cite{wikipedia} is a user-edited online encyclopedia.  The
system allows collaboration among its users to build its content. In most
cases, any user can change and update the contents of any article in the
encyclopedia; this policy is recently being revised with the rise in wikibots
that automatically inject spam content into wiki pages. The system maintains a
history of changes that allow any user to roll the article back to a previous
version, in case of unwanted additions.

\emph{Wayfinder}~\cite{wayfinder} is a P2P file sharing system that provides a
global namespace and automatic availability management. It allows any user to
modify any portion of the namespace by modifying, adding, and deleting files
and directories.  Wayfinder's global namespace is constructed by the system
automatically merging the local namespaces of individual nodes.
\emph{Farsite}~\cite{farsite} is a server less distributed file system. Farsite
logically functions as a centralized file server but its physical realization
is dispersed among a network of untrusted workstations.
\emph{OceanStore}~\cite{oceanstore} is a global persistent data store designed
to scale to billions of users.  It provides a consistent, highly-available, and
durable storage utility atop an infrastructure comprised of untrusted servers.
\emph{Cooperative File System}~\cite{CFS} is a global distributed Internet file
system that also focuses on scalability.  \emph{Ivy}~\cite{ivy} is a
distributed file system that focuses on allowing multiple concurrent writers to
files.

The work in~\cite{586138} considers a rating scheme using a distributed polling
algorithm.  These schemes and others like them, consider the files or resources
independently rather than within the context of a structure like a directory
structure and they do not permit users to choose among the best versions of a
given file. In~\cite{501175} the reputation of the rater is taken into account,
which is complementary to our contribution.

\section{Conclusion}

Most peer-to-peer file systems use keyword searches to discover files in the
network. Use of a directory network, where files are used as directory nodes,
is an emerging method for providing a browsable index of files. This approach
is difficult because of conflicts that occur when multiple users want to write
to the shared namespace. We overcome the problem by using a popularity based
system, where multiple versions (up to one version from each user) of a given
file or directory node are permitted and by default a user views the most
popular version of that node. Users may select a different version of the node
and the system keeps track of which which users are currently viewing which
nodes. We have built a prototype system, available online, which uses
BitTorrent and Kademlia. In this paper we showed the results of a comprehensive
simulation study of the ability for the popularity based system to promote high
quality files under a range of different user characteristics.

In our study we modeled the user characteristics and the resulting directory
structures that arise when a population of users behave in different ways. We
show that the popularity based system consistently gives rise to a default tree
that, while consisting of only a small fraction of all nodes in the system,
yields reasonably higher than average quality nodes. The popularity based
system is quite resistant to peer churn and can maintain quality with
reasonable frequency of user updates to nodes.

Broadly speaking, if users naturally select popular nodes over unpopular nodes
(with a probability proportional to the popularity) and choose to reselect if
the selected node is of low quality (with a probability proportional to the
quality) then the system allows for searches along paths that contain low
quality nodes and thus allows for discovery of high quality nodes further down
the tree. This is because low quality nodes can become popular just as fast
as high quality nodes, as users are unaware of quality until they view the node.
We could improve the simulation by improving the way in which quality
is assigned to nodes, e.g. quality may be averaged over updates and links to
high quality nodes could lead to increased quality, etc.

We have not yet considered the affects of attacks, such as collusion attacks
where a single user controls a number of peers and tries to promote the
popularity of low quality files. This is the focus of our future work.


\bibliographystyle{abbrv}
\bibliography{JacobsHarwood}


\end{document}